\documentclass{aa}  
\usepackage{graphicx}
\usepackage{gensymb}
\usepackage{txfonts}
\begin{document}
   \title{Average Fe K$\alpha$ emission from distant AGN\thanks{Based on observations obtained with {\it XMM-Newton}, an ESA science mission with instruments and contributions directly funded by ESA Member States and NASA.}}

   \author{A. Corral\inst{1},\inst{2}\and M.J. Page\inst{3}\and F.J. Carrera\inst{1}\and X. Barcons\inst{1}\and S. Mateos\inst{4}\and J. Ebrero\inst{1}\and M. Krumpe\inst{5}\and A. Schwope\inst{5}\and J.A. Tedds\inst{4}\and M.G. Watson\inst{4}}
   \offprints{A. Corral: corralra@ifca.unican.es}
   \institute{Instituto de F\'\i sica de Cantabria (CSIC-UC), 39005 Santander, Spain\\\email{corralra@ifca.unican.es}\and INAF - Osservatorio Astronomico di Brera, via Brera 28, 20121 Milan, Italy\\\email{amalia.corral@brera.inaf.it}\and Mullard Space Science Laboratory, University College London, Holmbury St. Mary, Dorking, Surrey RH5 6NT, UK \and X-ray \& Observational Astronomy Group, Department of Physics and Astronomy, University of Leicester, Leicester LE1 7RH, UK \and Astrophysikalisches Institut Potsdam, An der Sternwarte 16, D-14482 Postdam, Germany}

   \date{Received ,; accepted ,}

% \abstract{}{}{}{}{} 
% 5 {} token are mandatory
 
  \abstract
{One of the most important parameters in the XRB ({\it X-ray
    background}) synthesis models is the average efficiency of
  accretion onto SMBH ({\it super-massive black holes}). This can be
  inferred from the shape of broad relativistic Fe lines seen in X-ray
  spectra of AGN ({\it active galactic nuclei}). Several studies have
  tried to measure the mean Fe emission properties of AGN at different
  depths with very different results.}
{We compute the mean Fe emission from a large and representative
  sample of AGN X-ray spectra up to redshift $\sim$ 3.5. }
{We developed a method of computing the
rest-frame X-ray average spectrum and applied it to a large sample
(more than 600 objects) of type 1 AGN from two complementary medium
sensitivity surveys based on XMM-Newton data, the AXIS and XWAS
samples. This method makes use of medium-to-low quality spectra
without needing to fit complex models to the individual spectra but
with computing a mean spectrum for the whole sample. Extensive quality
tests were performed by comparing real to simulated data, and a
significance for the detection of any feature over an underlying
continuum was derived.}
{We detect with a 99.9$\%$ significance an unresolved Fe K$\alpha$ emission line around 6.4~keV with an EW $\sim$ 90 eV, but we find no compelling evidence of any significant broad relativistic emission line in the final average spectrum. Deviations from a power law around the narrow line are best represented by a reflection component arising from cold or low-ionization material. We estimate an upper limit for the EW of any relativistic line of 400 eV at a 3$\sigma$ confidence level. We also marginally detect the so-called Iwasawa-Taniguchi effect on the EW for the unresolved emission line, which appears weaker for higher luminosity AGN.}
{We computed an upper limit for the average relativistic Fe K$\alpha$ line contribution that is significantly lower than previously reported values from similar analyses. Our results, however, are in excellent agreement with individual analyses of local AGN samples. We attribute this difference either to our more sophisticated method of modeling the underlying continuum, to intrinsic differences in source populations, and/or to the uneven data quality of the individual spectra of the various samples.}


   \maketitle
%
%________________________________________________________________

\section{Introduction}

The {\it X-ray background} (XRB) has been almost completely resolved
into discrete sources at energies up to several~keV, using X-ray
surveys at different depths (see Gilli et al. \cite{gilli} for a
recent review). The vast majority of the sources that comprise the XRB
at these energies are {\it active galactic nuclei} (AGN). The
integrated emission of AGN reflect the history of accretion onto {\it
  super-massive black holes} (SMBH) over cosmic time. In order to
reproduce the shape and intensity of the XRB, synthesis models are
constructed by using a mixture of AGN spectra. The most important
physical parameters in these models are the AGN intrinsic column
density and accretion-rate distribution and their evolution as a
function of luminosity and redshift, as well as the average radiative
efficiency of accretion onto SMBH. Since AGN emission comes from
accretion onto SMBH, by comparing the energy radiated by distant AGN
to the mean mass in local SMBH (Soltan \cite{soltan}), an estimate of
the efficiency of AGN in converting mass to energy -- the mean
radiative efficiency -- can be computed. Following this argument, the
mean radiative efficiency has been estimated to range from 10 to 15 \%
(even more than 20\% for the most massive systems) from recent studies
(Elvis et al., \cite{elvis}; Yu \& Tremaine \cite{yu}; Marconi et al.,
\cite{marconi}). Although this method suffers from several
uncertainties, they can only lead to an underestimate of the radiative
efficiency. The accretion efficiency depends on the orbit where most
of the accretion material is located, the {\it innermost stable
  circular orbit} (ISCO). The ISCO, in turn, is a function of the SMBH
spin. For a non-rotating Schwarzschild BH the efficiency is
constrained to be below 6\% (Lyden-Bell, \cite{lyden}). To achieve an
efficiency stronger than this requires a rapidly rotating SMBH
(Thorne, \cite{thorne}).\\

The mean accretion efficiency can be inferred from the relativistic
profile of an emission line emitted close to the SMBH. Emission lines
are usually seen in the X-ray spectrum of AGN, the Fe K$\alpha$ line
(6.4-6.9~keV depending on ionization state) being the best studied
one. If the line is emitted close enough to the SMBH, it shows a broad
relativistic profile which is more pronounced for higher SMBH spin due
to the ISCO becoming smaller and therefore gravitational redshift (and
other General Relativistic effects) being larger (Fabian et al.,
\cite{fabian89}; Laor \cite{laor}).\\

Early results from the ASCA era suggested that broad relativistic
lines might be common in type 1 AGN. Surprisingly, however, they have
been significantly detected and characterized in only a small number
of sources (Reynolds \& Nowak, \cite{reynolds}), MCG-6-30-15 being the
best studied one (Tanaka et al., \cite{tanaka}; Fabian \& Vaughan,
\cite{fabian03}). The number of counts collected in the AGN X-ray
spectra turns out to be the limiting factor, since very accurate
modeling of the continuum below the line is critical to properly
measure the line properties (Guainazzi et al. \cite{gua}). Another
possibility, somewhat related to the previous one, is that the extreme
relativistic blurring along with high inclination angles can both
widen and weaken the line making it undetectable (Fabian \& Miniutti
\cite{fabian05}). The direct average Fe line contribution to the XRB
is therefore still unknown although it has been estimated to range
from 3$\%$ to 7$\%$ (Gilli et al., \cite{gilli99}; Gandhi \& Fabian
\cite{gandhi}) depending on the assumptions made about the shape and
     {\it equivalent width} (EW). These assumptions affect the
     spectral templates used to compute the XRB, thus affecting its
     shape.\\

Several studies have been performed over samples of local AGN in order
to characterize Fe K$\alpha$ emission in the local Universe. Nandra et
al. (\cite{nandra}) performed a spectral analysis of a sample of 26
type 1 to 1.9 Seyferts galaxies (z $<$ 0.05) observed by {\it
  XMM-Newton}. They found that a relativistic line is significantly
detected in half of their sample (54$\pm$10 per cent) with a mean EW
of $\sim$~80~eV, when fitted as a broad Gaussian. Guainazzi et
al. (\cite{gua}) and de La Calle et al. (in preparation) carried out a
similar analysis over a larger sample of local type 1 and 2
radio-quiet AGN, excluding highly obscured type 2 sources. They
detected relativistic lines in 25\% (50\% for well-exposed sources) of
their sample with an EW $\sim$ 200 eV. Both studies, although
computing a similar average EW, found a high dispersion in the
individual values. Nevertheless, this kind of analysis cannot be
extended to higher redshifts due to the more limited quality of the
spectra of distant AGN. It is, however, at high redshift where the
measurement of the accretion efficiency is most relevant to the
synthesis of the XRB. \\

Improving the {\it signal to noise ratio} (SNR) by averaging as many
AGN spectra together as possible is the best solution since spectra of
the same type of AGN are expected to display similar spectral
characteristics. Recent studies have constructed X-ray averaged
spectra for AGN but obtained differing results. In Streblyanska et
al. (\cite{streb}) (hereafter S05) the results of this kind of
analysis are presented for 53 type 1 and 41 type 2 AGN in the pencil
beam Lockman Hole {\it XMM-Newton} field (z $<$ 4.5). They found a
broad relativistic line in the final averaged spectra with an EW of
400 and 600 eV for type 2 and type 1 AGN, respectively. Brusa, Gilli
\& Comastri (\cite{brusa}) found qualitatively similar results when
stacking the AGN spectra contained in the 1Ms Chandra Deep Field South
in 7 different redshift bins from z =0.5 to 4.0, the broad line EW
being slightly weaker than in S05 but consistent within
errors. Longinotti et al.(in preparation), stacked X-ray spectra from
a local sample of AGN (extended from the sample of Guainazzi et al.,
\cite{gua}), and found that the EW for the broad relativistic
contribution is never larger than 80 eV, either when stacking the
whole sample or different sub-samples. Given this apparent divergence
of results at high and low redshifts, it is vital to consider
carefully and refine the averaging or stacking method applied to faint
object spectra.\\

We present here a new method to construct a rest-frame X-ray averaged
spectrum for a large sample of AGN at different redshifts, in order to
characterize the mean Fe K$\alpha$ emission. We compiled a large
sample of spectroscopically identified AGN from two wide angle and
large but complementary medium sensitivity surveys: AXIS ({\it An
  International XMM-Newton Survey}; Carrera et al., \cite{carrera}) in
the Northern Hemisphere and XWAS ({\it XMM-Newton Wide Angle Survey};
Tedds et al., in preparation) in the South in order to maximize the
total number of counts. Since our sample is composed of low to medium
quality X-ray spectra, we developed a method that takes into account
the effects of the continuum emission around the emission line and the
counting statistics, without needing to fit complex models to the
individual spectra, but modelling the underlying continuum using
simulations. We have extensively tested this method and a significance
for the emission line detection can be derived from it.  Some results
using this method were presented by us for a smaller sample of type 2
QSOs in Krumpe et al. (\cite{krumpe}), hereafter K08, showing a strong
FeK$\alpha$ emission line in the final averaged spectrum but only for
the low luminosity sources.\\ The paper is organized as follows: in
sections 2 and 3 we describe the spectral extraction and the averaging
method, respectively. In section 4 we discuss the most important
results for the total sample and different sub-samples. Conclusions
are presented in Section 5.  All the quoted errors are at the
1$\sigma$ confidence level throughout the paper. We assume $H_{0}$=70
$km\,s^{-1}\,Mpc^{-1}$, $\Omega_{M}$=0.3 and $\Omega_{\Lambda}$=0.7.

\section{X-ray data}
We extracted the spectra for the XMS ({\it XMM-Newton Medium Survey})
fields within the AXIS sample ($\sim$ 40 {\it XMM-Newton}
observations) and the complete XWAS sample ($\sim$ 150 {\it
  XMM-Newton} observations), but only included in our analysis those
AGN that are optically identified as type 1 AGN in these fields, i.e.,
those AGN that show broad emission lines (velocity widths $\geq$ 1500
$km s^{-1}$) in their optical spectra. We selected only type 1 AGN due
to the larger number of sources and the better quality X-ray spectra
for this type of object in both samples. Complete descriptions of both
samples can be found in Carrera et al. (\cite{carrera}) and Tedds et
al. (in preparation) for the AXIS and XWAS samples, respectively.\\
 
We processed the refinalized ODFs ({\it observation data files}) from
the {\it XMM-Newton} archive using the {\it Science Analysis Software}
(SAS\footnote{http://xmm2.esac.esa.int/sas/8.0.0/}) version 6.1.0 for
the AXIS sample, and 6.5.0 for the XWAS sample, the latest software
and calibration available at the time of our spectral extraction
(2006). We processed the ODFs in order to obtain the calibrated images
and event lists using the SAS pipeline chains {\tt emchain} and {\tt
  epchain} for the EPIC-MOS and EPIC-pn data, respectively.\\

After filtering out the event lists for high background intervals, we
extracted the source spectra in circular regions maximizing the SNR
for all detectors via the SAS task {\tt eregionanalyse}. The
background spectra were taken in annular regions centered on the
source position. If any other source falls within this region, we
excluded the new source region from the background region. If the
resulting background region was too small or fell near bright sources,
we substituted it by a circular nearby source-free region. We
  selected single, double, triple and quadruple events in the MOS1 and
  MOS2 (pattern $\leq$ 12) and single and double events in the pn
  (pattern $\leq$ 4). For the pn we selected only high quality events
  ({\tt FLAG==0}). We merged MOS1 and MOS2 spectra corresponding
  to the same source, observation and filter. We also obtained
response matrices for each individual source, camera and observation
via the SAS tasks {\tt rmfgen} and {\tt arfgen}.\\

In some cases, there is more than one observation of the same
source. Since we do not expect individual source variability to
significantly affect our results, we improved the spectral quality by
merging the spectra for the same source, observed with the same EPIC
camera, at a similar off-axis angle on the same CCD chip and with the
same filter. If filter, CCD chip or off-axis were different the
spectra were kept separate. The resulting individual pn and MOS
spectra, even if corresponding to the same source, were treated as
separate contributions to the final average spectrum. \\

\begin{figure}
\centering
\includegraphics[width=6cm,angle=-90]{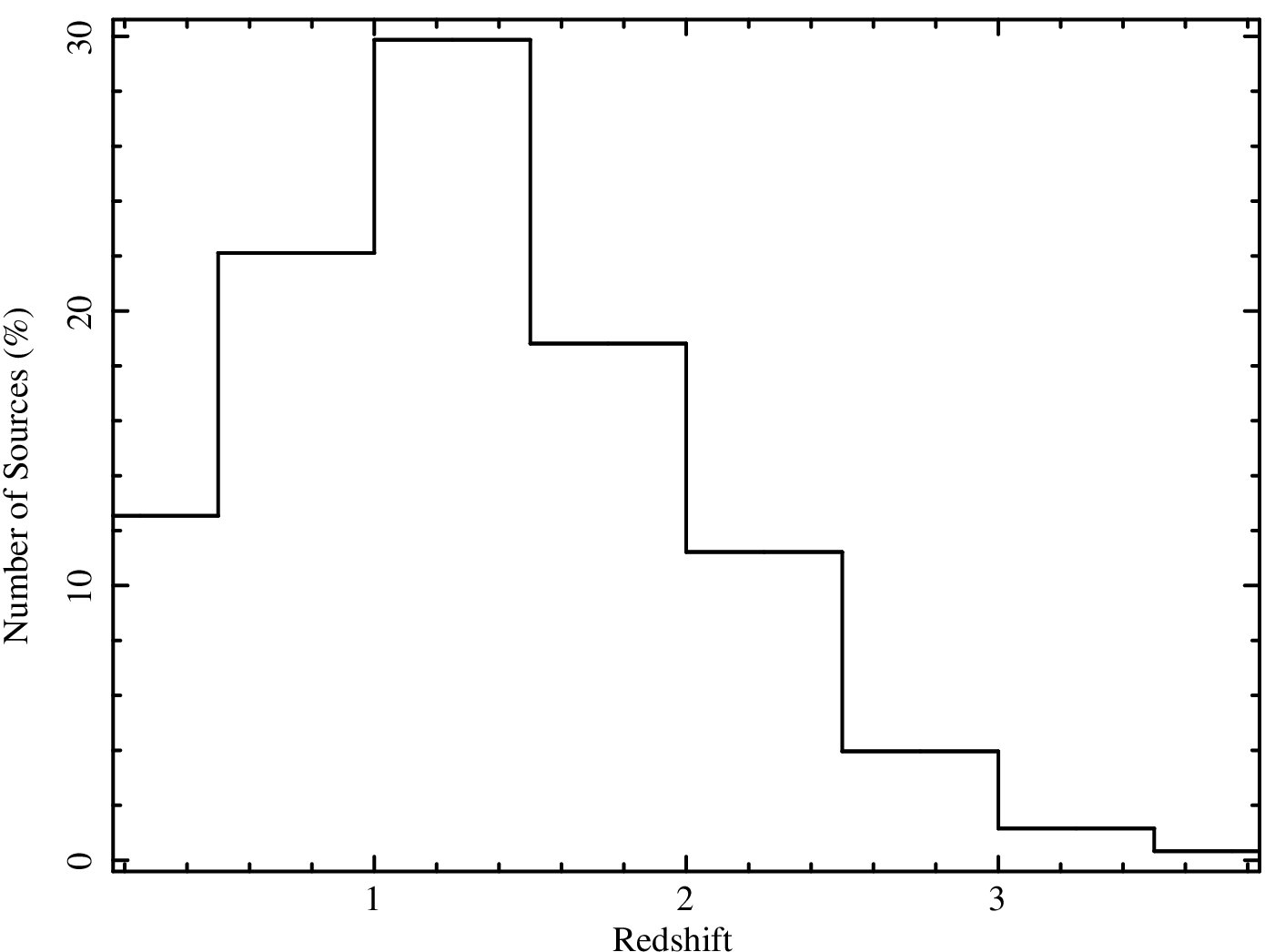}
\includegraphics[width=6cm,angle=-90]{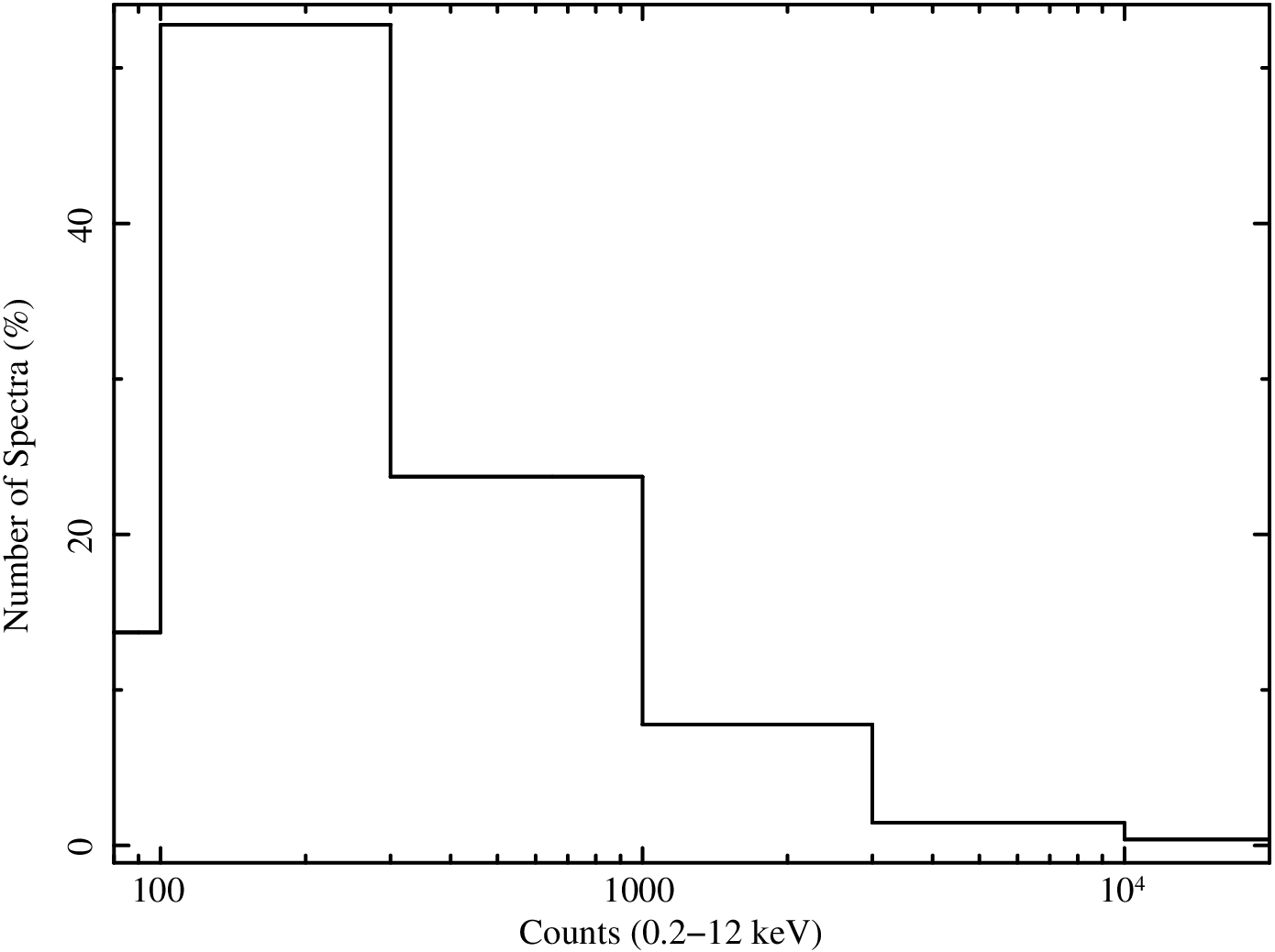}
\includegraphics[width=6cm,angle=-90]{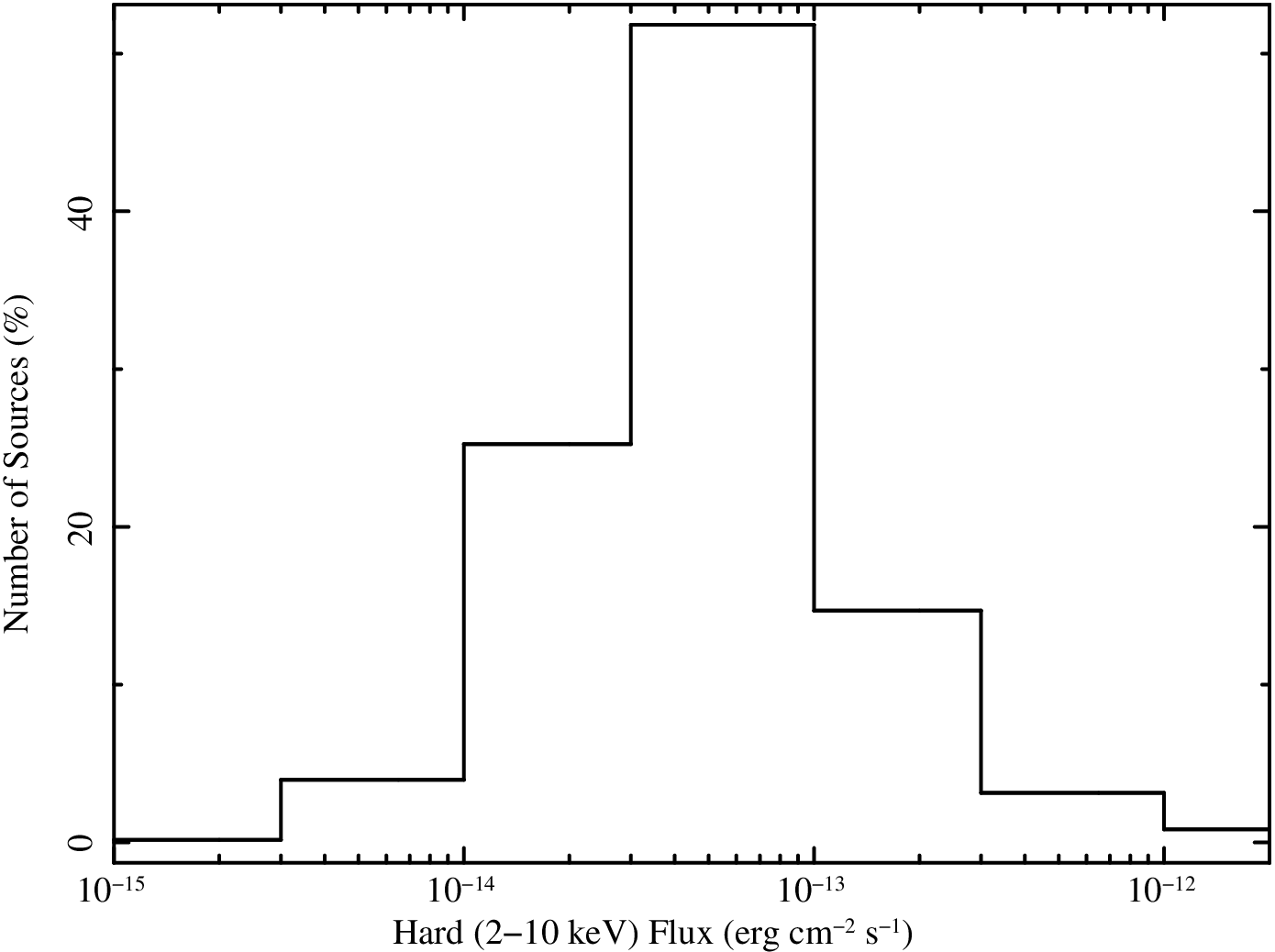}
\caption{{\it Top panel}: Redshift distribution. {\it Central panel}: Counts distribution in the total band (0.2-12.0~keV). {\it Bottom panel}: Flux distribution in the standard hard band (2.0-10.0~keV).}
\label{distbs}
\end{figure}

After the spectral extraction and merging procedure, we selected only
the spectra (pn or MOS) containing more than 80 counts in the total
band (0.2-12 keV), in order to preserve a minimum spectral quality in
the sample. This provided a sample of 606 optically identified type 1
AGN corresponding to more than 1000 individual pn and MOS spectra. Out
of these 606 sources, 488 have been observed only once, 78 twice, 16
three times, 6 four times and 18 five times. Note that, having more
than one observation does not mean having more than one spectrum,
since the final number of spectra used depends on the merging
procedure. The counts and redshift distributions, along with the flux
distribution in the standard hard (2-10 keV) band, are shown in
Fig.~\ref{distbs}. It should be noted that we did not intend to
construct a complete or flux-limited sample in any way, but to collect
as many counts as possible in order to obtain a high SNR averaged
spectrum. Therefore the sample comprises objects having a wide range
of X-ray luminosities and redshifts. Fluxes are, however, mostly
concentrated in the 10$^{-14}$-10$^{-13}$ erg cm$^{-2}$ s$^{-1}$ flux
range, significantly (but not largely) brighter than those in the
Lockman Hole sample studied in S05. Luminosities, however, are higher
in average for the latter sample due to the redshift distribution
peaking close to z=2, whereas it peaks closer to z=1 in our sample. In
the stacking analysis presented in K08 all the available spectra are
considered in the averaging procedure, no matter how many counts they
have. Given that our sample is much larger than the one in K08, we can
thus place a much more limiting constraint on the minimum spectral
counts threshold so as to improve the quality of our final averaged
spectrum.  \\
\section{Averaging Method}
Having extracted source and background spectra and generated response
matrices for each source we next needed to unfold
the spectra, i.e. to recover the original source spectra before detection 
prior to combining the intrinsic spectra from different sources.\\
\subsection{Unfolding the spectra}
To obtain the unfolded spectrum for each source, we used {\tt XSPEC
  v12.4} (Arnaud et al., \cite{arnaud}). We first grouped the spectra
to have at least 10 counts per bin (the minimum number of counts
required to use the $\chi^2$ minimization technique). We then used a
single power law model corrected by photoelectric absorption, fixed at
the Galactic value for each source position, and intrinsic
photoelectric absorption at the source redshift. It should be noted
that our intention was not to obtain the best fit parameters but a
model good enough to unfold the spectrum without severe biases {\it
  and} to obtain simulated data that represent the actual shape of the
spectra (see Appendix A). In fact the small number of spectral counts
for most of the sources did not allow us to fit a more complex model
to the data. To avoid the contribution of any soft excesses we fitted
each individual spectrum above 1~keV (rest-frame) leaving the
intrinsic hydrogen column density, power law slope and normalization
as free parameters. Then, we applied this ``best fit model'' to the
ungrouped spectrum, obtaining the unfolded spectrum in physical units
(keV cm$^{-2}$ s$^{-1}$ keV$^{-1}$) and free of instrumental effects
({\tt eufspec} command in {\tt XSPEC}). Note that given the small
number of counts in the spectra presented in K08, their spectral fit
was performed over the full energy range (0.2 to 12 keV) and the
spectral slope was fixed to 1.9. It is known that performing the
spectral fit in this way might introduce some biases, mostly
significant at low energies, that can affect the continuum shape due
to the intrinsic absorption although not significantly affecting
narrow features.  \\

Below 3 keV, the unfolding process is highly dependent on the model
used, due to the shape of the effective area and the limited spectral
resolution (FWHM $\sim$ 80 and 150 eV at 1 and 6.4 keV, respectively),
meaning that caution must be applied in deriving any result in our
analysis of the final average spectrum below that energy. Narrow
(unresolved) features, both in the model or in the data itself, could
also affect the unfolding process above these energies. A narrow
emission (or absorption) line, for example, will be widened during the
unfolding process if it is not included in the fitting model. Although
less significant than the softer ones, these features should be taken
into account very carefully, especially around the position where we
expect the Fe emission line, as they can distort its shape. Absorption
edges are not expected to significantly affect the spectral shape at
these energies due to the limited amount of absorption for type 1
AGN. Given the quality of our spectra, we cannot fit these features
directly from the individual spectra, so we estimate their
contribution to the continuum around the emission line position by
using simulations, as we explain at the end of this section. Plots
showing the shape of the effective area and the effect of the
absorption over a power law can be seen in Appendix A. \\

Once all the spectra are in physical units we corrected them for
Galactic absorption (i. e., we de-absorb the spectrum) via a table
model extracted from the {\tt phabs XSPEC} model, and shift them to
their rest frame using the redshifts derived from the optical
identifications.\\
\subsection{Rescaling the spectra}
We require that each spectrum contributes with the same weight to the
final averaged spectrum and the simplest way to achieve that, without
distorting the individual spectral shape, is by dividing each spectrum
by a certain value so that every rescaled spectrum has the same
flux. Since the differences in spectral shapes could be important, the
first issue we have to solve is to select a spectral range to achieve
a rescaling not affected by spectral features. It is clear that we
must exclude the energy region where the Fe K$\alpha$ line is expected
to fall but also exclude lower energy regions due to absorption
effects. We also cannot include higher energies to avoid the
contribution of a spectral band where large errors are common. After
testing several energy bands within the previous constraints and using
simulated data, we selected the 2-5~keV band to rescale the
spectra. Tests with our simulations showed that this band recovers the
input spectral shape in the most accurate way while minimizing the
errors.  We then proceeded by computing the fluxes in the 2-5~keV
rest-frame band for each spectrum and dividing each spectrum by its
corresponding value so that all the rescaled spectra have the same
2-5~keV rest frame flux in units of keV cm$^{-2}$ s$^{-1}$. \\

Although now rescaled, each spectrum was expressed on a different
energy grid because of the different channel sizes at different
energies and shifting to rest-frame. In order to achieve some
uniformity in the errors of the average spectrum across the whole
energy band, we constructed a new energy grid for the final averaged
rest-frame spectrum ensuring a minimum number of real source counts
($\sim$ 1000) in each new bin. To select these new bins, we used the
individual source spectra as measured in counts.  We first shifted
these spectra in counts to the rest frame, rebinned them to a common
energy grid composed of narrow bins (bin widths $\sim$ 40 eV) and
added them all together. We then grouped the narrow bins so as to
distribute the counts in the most uniform way and so that each new bin
contains at least 1000 real source counts. We finally distributed the
rescaled flux density values for each individual spectrum among the
new energy bins in the following way:\\
\begin{equation}
 S_j^{'} = \sum_{i \subset j} {{S_i \Delta{\epsilon_i}f_{ij}} \over {\Delta^{'}{\epsilon_{j}}}}
\end{equation}
\begin{equation}
 f_{ij} = {{min(\epsilon_{imax},\epsilon^{'}_{jmax}) - max(\epsilon_{imin},\epsilon^{'}_{jmin})} \over {\Delta{\epsilon_i}}}
\end{equation}
Where $S_j^{'}$, $\Delta^{'}{\epsilon_j}$ and $S_i$,
$\Delta{\epsilon_i}$ are the flux density values (in keV cm$^{-2}$
s$^{-1}$ keV$^{-1}$) and widths (in keV) of the new and old bins,
respectively, and $f_{ij}$ represents the fraction of the old bin {\it i}
that covers the new bin {\it j}.
\subsection{Averaging the spectra}
Once we have the rest frame rescaled and rebinned spectra, we
simply averaged them using an un-weighted standard mean. Because of the
quite large dispersion in the redshift distribution, high energies are
only significantly detected for a few objects. We did not take into 
account those spectral ranges that were not significantly represented 
by at least 10 contributing spectra. The individual errors were propagated 
as Gaussian during the entire process, the final errors being computed in 
the following way:

\begin{equation}
 Error_j ={ \sqrt{{\sum_{i \subset j}^{N}} {{\sigma_i^{2}}}} \over {N}}
\end{equation}
Where $\sigma_{i}$ is the individual error corresponding to the spectrum
{\it i} at the bin {\it j} and N is the number of spectra contributing
to bin {\it j}.\\

\begin{figure}
\centering
\includegraphics[width=6cm,angle=-90]{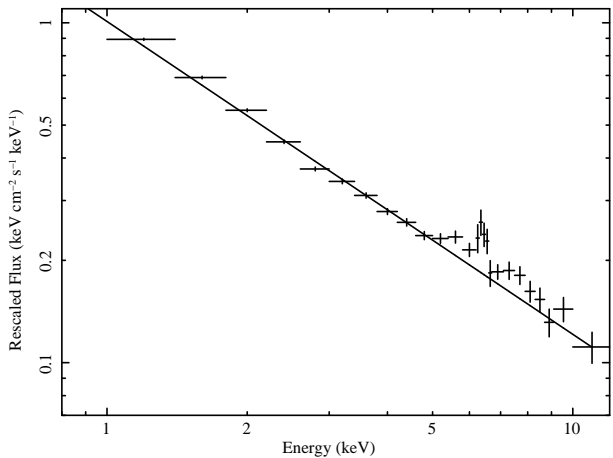}
\caption{Averaged spectrum along with the power law fit.}
\label{average}
\end{figure}

The resulting averaged spectrum is shown in Fig.~\ref{average}. We fit
a power law between 2 and 10~keV, excluding the 4-7~keV range,
obtaining a value of $\Gamma=1.92\pm0.02$. We clearly see a narrow
emission feature around 6.4~keV but also a broad excess of emission
from 5 to 10 keV. Instead of simply fitting a model to the final
average spectrum and in order to derive a significance in the
detection of the observed features, we decided to construct an
underlying continuum by using simulations. In this manner we could 
quantify how the results are affected by counting statistics, spectral 
features due to photoelectric absorption and the averaging method itself. 
An extensive discussion about the simulations that have been carried 
out is presented in Appendix A.\\

\subsection{Constructing a continuum}
In order to construct the underlying continuum against which we could
search for any Fe emission or other spectral components, we simulated
each individual source and background spectrum by using the {\tt
  XSPEC} command {\tt fakeit}. Each spectrum was simulated 100 times,
including Poisson counting noise and keeping the same 2 to 8~keV
observed flux, exposure time and calibration matrices as in the real
data. We used the ``best-fit model'' (absorbed power-law) obtained when
fitting the individual spectra and we applied to the simulated spectra
exactly the same averaging method as for the real data. By simulating
100 times each source we were able to reduce spurious features as well
as to construct 100 absorbed power-law simulated continua. From these
100 simulated continua we could estimate the significance in the
detection for any feature in the real averaged spectrum. By removing
the 32 and 5 extreme values for each simulated continuum at each
energy bin, we computed the normalized flux intervals that encompass
68\% and 95\% of the simulated values. Therefore, a spectral excursion
above or below these limits is detected at the 1$\sigma$ or 2$\sigma$
level, respectively. Other spectral components or features,
not taken into account in our simulations could contribute to the
underlying continuum in the real spectra, but the construction of an
underlying continuum in this way allowed us to build a baseline
spectrum that accounted for the absorbed power-law and for all the
effects introduced by the averaging process.  \\

\begin{figure}
\centering
\includegraphics[width=6cm,angle=-90]{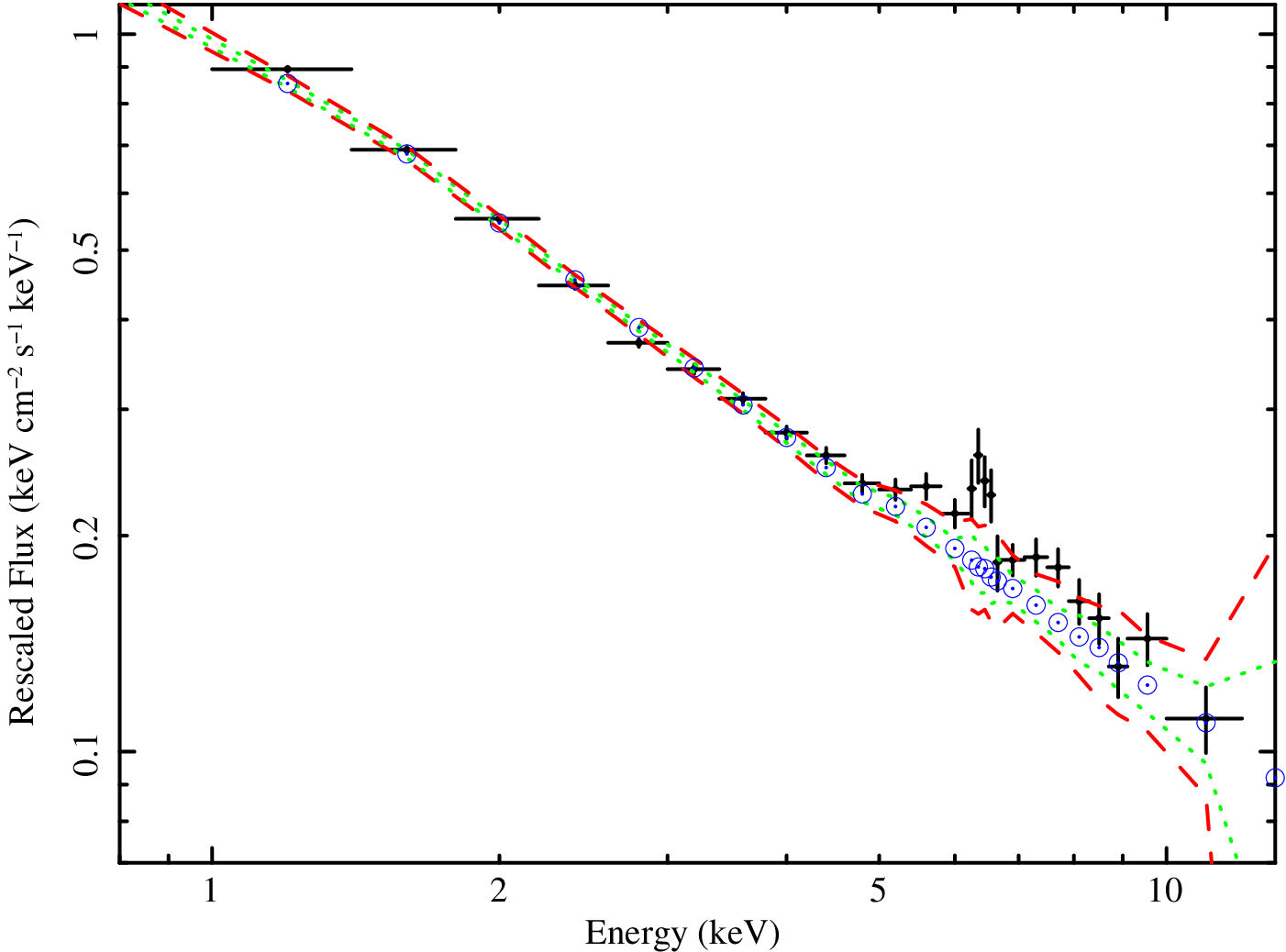}
\includegraphics[width=6cm,angle=-90]{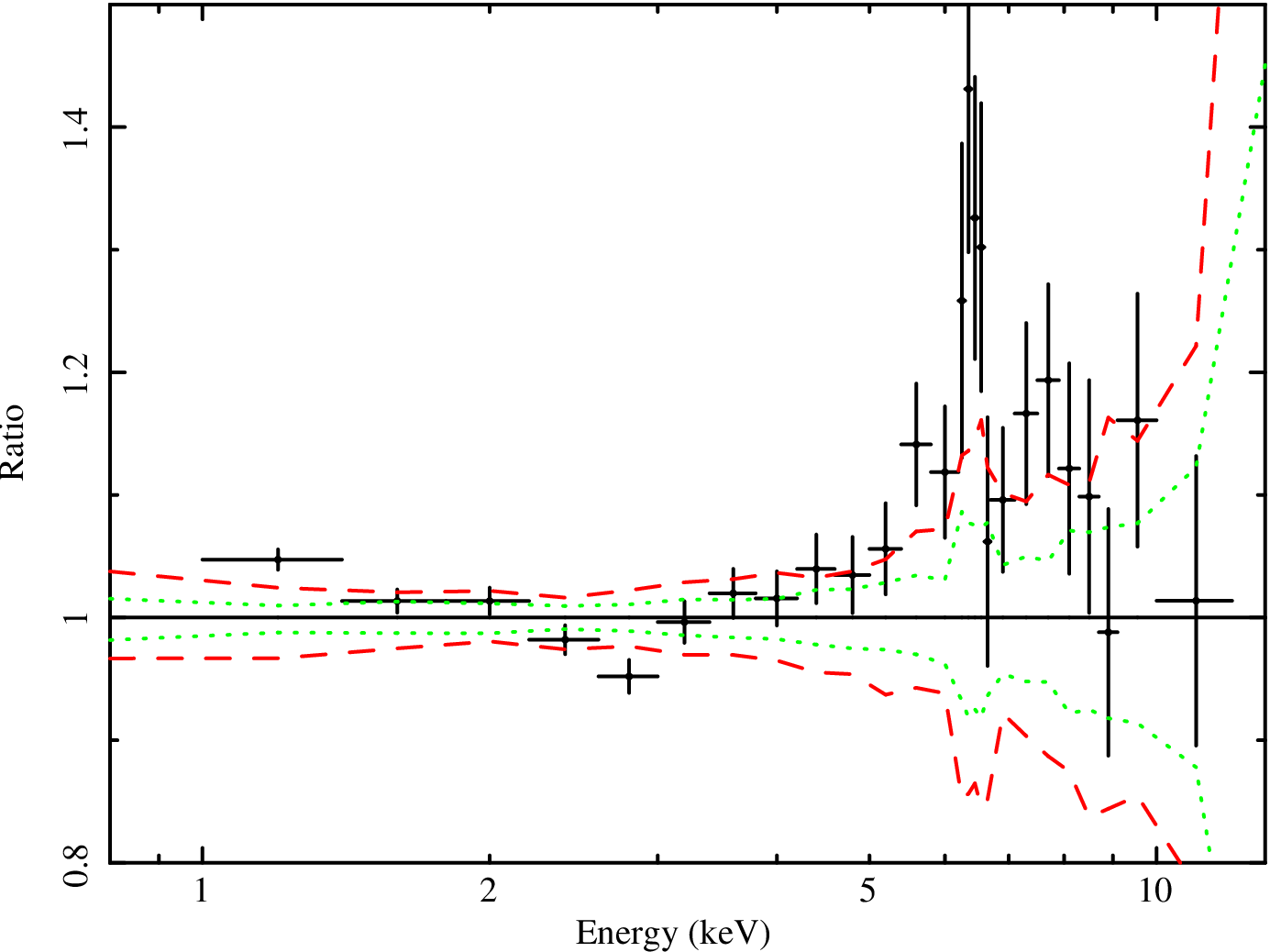}
\caption{ {\it Top panel}: Averaged spectrum (error bars) along with the
  simulated continuum (open circles). Dotted and dashed lines
  represent the 1$\sigma$ and 2$\sigma$ continuum confidence limits,
  respectively. {\it Bottom panel}: Averaged spectrum to simulated
  continuum ratio and confidence limits.}
\label{spectrum}
\end{figure}

The averaged spectrum, simulated continuum and 1$\sigma$ and 2$\sigma$
limits in each bin are shown in Fig.~\ref{spectrum}. We clearly see
emission features between 5 and 10~keV but only the narrow peak is 
clearly above the 2$\sigma$ contour. We also see that some apparent
deviations from a power law shape present in the averaged
spectrum are also present in the simulated continuum. This result
emphasizes the importance of properly computing the underlying
continuum in order to fit the emission line.\\

In order to check if a number of ``extreme'' sources are responsible
for the observed emission features, we performed a safety test. We
removed in each energy bin all the spectra that deviate more than 3
times the standard deviation from the average value. The resulting
averaged spectrum shows the same overall shape as for the whole
sample, the 5 to 10 keV emission residuals being smaller than the ones
in Fig.~\ref{spectrum}, but consistent within errors. This could be
due either to the fact that we are actually removing the sources that
show more prominent emission in that range, or we are simply reducing
the noise by removing the lowest quality spectra. In any case, the
differences found are not significant.

\section{Discussion and Results}
\begin{table*}[t]
\centering
\caption[]{Best fit parameters.}
$$
\begin{array}{l c c c c c c c c} 
\hline
{\rm Model} & {\rm E_{gaus}} & {\rm \sigma_{gaus}} & {\rm EW_{gaus}} & {\rm Reflection\;Factor}& {\rm Inclination\;Angle} & {\rm Ionization\;Parameter} &{\rm EW_{laor}} & \chi^{2}/d.o.f\\
 & {\rm (keV)} & {\rm (eV)} & {\rm (eV)} & & {\rm (\degree)} & {\rm (erg\,cm\,s^{-1})}  & {\rm (eV)} & \\
\hline 
\hline 
{\rm Baseline\;model} & {\rm 6.34^{n}} & {\rm 200^{p}} & {\rm 200^{n}} & \dots & \dots & \dots & \dots & {\rm 41/18}\\
& & & & & & & & \\
{\rm Baseline\;model\;+\;pexrav} & {\rm 6.36^{+0.05}_{-0.05}} & {\rm 80^{+60}_{-80}} & {\rm 90^{+30}_{-30}} & {\rm 0.50^{+0.15}_{-0.20}} & {\rm 60^{+15}_{-20}} & \dots & \dots &{\rm 14/15} \\
& & & & & & & & \\
{\rm Baseline\;model\;+\;kdblur(reflion)} & {\rm 6.37^{+0.04}_{-0.04}} & {\rm 80^{+40}_{-40}} & {\rm 80^{+20}_{-20}} & \dots & {\rm 30^{f}} & {\rm \xi<30} & \dots & {\rm 15/16} \\
& & & & & & & & \\
{\rm Baseline\;model\;+\;laor} & {\rm 6.37^{+0.06}_{-0.06}} & {\rm 10^{+10}_{-10}} & {\rm 50^{+30}_{-20}} & \dots & \dots & \dots & {\rm 350^{+70}_{-90}} & {\rm 26/17} \\
\noalign{\smallskip} 
\hline
\end{array}
$$
\begin{list}{}{}
\item Model parameters not displayed correspond to default values.
\item[$^{n}$] poor fit does not allow error calculation. 
\item[$^{p}$] fit parameter pegged at hard limit.
\item[$^{f}$] fixed parameter. 
\end{list}
\label{models}
\end{table*}
Assuming the simulated continuum represents the actual underlying
continuum below the emission features, we measured a total excess
emission with EW of $\sim$~600~eV, between 5 and
10~keV. Obviously, this value may correspond to the addition of
multiple components in this range. As can be seen in
Fig.~\ref{spectrum}, a relativistic profile is not clearly present in
the resulting averaged spectrum. However, a mixture of different
relativistic profiles at different disk inclinations and innermost
radii is not expected to result in a straightforward relativistic
profile. In the same way, a mixture of warm absorbers or reflection
components should not be perfectly fitted by a one-component model of
any of these kind of spectral features.\\

We performed a spectral analysis of the averaged spectrum using {\tt
  XSPEC v12.4}. We excluded energies above 15~keV where the errors
were very large and also energies below 3~keV to avoid the
contribution of soft features that can be due to the unfolding
procedure. As mentioned in Sec. 3.1, the spectral shape below 3 keV is
highly dependent on the model used to unfold the spectra. Therefore,
the apparent ``soft-excess'' below 2 keV and the ``absorption''
feature around 2.5 keV could be due to the averaging process and we
should prevent ourselves from assigning them any physical meaning. In
any case, the ``soft-excess'' accounts only for less than 5 $\%$ of
the soft flux, and the ``absorption'' feature is barely significant.

We defined a ``baseline model'' composed of the following two
components:\\

-- A table model computed from the simulated continuum, with no free
parameters. As explained above, this should account for an absorbed
power law. Leaving the continuum's normalization to vary does not
significantly change the results, so we fixed it to prevent an 
unphysical broad line fitting multiple spectral bumps.  \\

-- A narrow Gaussian line with energy and normalization as free
parameters. The line width, also a free parameter, is constrained to
be below 200 eV. Given the widening due to EPIC response and the
averaging process, $\geq$~100 eV, this limit ensures a narrow line
fit. 
\\

Using this baseline model only we obtained a poor fit (41/18
$\chi^{2}$/d.o.f), but the narrow line is detected at $>3\sigma$
significance level ($\Delta\chi^{2}\,>$ 14.16 for 3 additional
parameters) with an EW~$\sim$~200 eV. The line is centered around 6.4
keV so it corresponds to neutral or low-ionization Fe K$\alpha$, i.e
it likely comes from distant and cold material such as the putative
torus in the Unified Schemes (Antonucci \cite{anto}). Observations
have shown that a narrow emission line corresponding to Fe K$\alpha$
is almost an ubiquitous characteristic in local AGN X-ray spectra (Page
et al., \cite{page}; Nandra et al., \cite{nandra}). We confirm and
extend this important result to the distant Universe.\\

It has been claimed that complex absorption and high-density ionized
absorbers can mimic the red tail of a broad relativistic line (Reeves
et al., \cite{reeves}; Turner et al., \cite{turner}). However, ionized
absorbers are often observed in AGN and we find that adding cold
(partially or totally covering the primary source) or warm ({\tt
  absori}, Zdziarski et al. \cite{zdz}) absorption does not improve the
fit at all.\\

The next step was to fit a reflected component from neutral material
({\tt pexrav} model, Magdziarz \& Zdziarski \cite{mad}). We obtain an
improvement on the fit of $>$3$\sigma$ significance level
($\chi^{2}$/d.o.f = 14/15), for a reflection fraction R $\simeq$~0.5
($\Omega/2\pi$ $\sim$~0.5) and inclination angle i
$\simeq$~60~$\degree$, assuming solar Fe abundance. It should be noted
that we fitted a reflection model for a given inclination angle to a
spectrum that corresponds to a mixture of different reflection
components with different inclination angles. Unfortunately, the
quality of the averaged spectrum does not allow us to recover the
inclination angle distribution for the sample so as to obtain an
angle-average model. Leaving the Fe abundance free only results in a
slightly higher value for it and no significant improvement of the
$\chi^{2}$. Adding relativistic blurring to the reflected component
({\tt kdblur} convolution model using a {\tt laor} profile) did not
improve the fit. Note that the amount of reflection along with the
measured narrow Fe K$\alpha$ EW are consistent with reflection from
distant Compton-thick matter, such as the torus, for type 1 AGN
(Reeves et al., \cite{reeves03}). \\

We also tried to fit a reflected component from an ionized disk ({\tt
  reflion} model, Ross \& Fabian \cite{ross}). We obtained quite a
good fit ($\chi^{2}$/d.o.f = 19/16), slightly worse than the previous
one, the disk being in a low-ionized state. Modifying it by
relativistic blurring, we obtained a similar goodness of fit
($\chi^{2}$/d.o.f = 15/16, all {\tt kdblur} parameters fixed to their
default values) as derived for the neutral reflection. However,
studying this model in detail, we found it provides only an upper
limit for the ionization parameter ($\xi<$30, neutral to low
ionization disk), that pegs at the lower limit permitted by the
model. Moreover, the effect of the relativistic blurring is to smear
the emission lines to the point that the resulting shape is almost
identical to the {\tt pexrav} model. Besides that, the fit does not
depend on the remaining disk parameters and, allowing them to vary,
results in unreasonable values, like an inclination angle for the disk
higher than 60~$\degree$. Therefore, we cannot distinguish (in terms
of $\Delta\chi^{2}$) between neutral reflection from distant material
and ionized reflection from a low-ionization accretion disk. However,
the computed parameters for the neutral reflection component, as long
as the central energy for the narrow Fe K$\alpha$ line and its EW
(6.36 keV and 90 eV, respectively), favor the neutral reflection
scenario rather than the ionized one (Reeves et al., \cite{reeves01};
George \& Fabian, \cite{george}). Furthermore, the relativistically
blurred low-ionization disk reflection just seems to be mimicking the
neutral reflection shape instead of actually fitting relativistic
effects.  \\

As a safety test, we checked if the measured reflected component could
be due to the dispersion in the power law indices, since a mixture of
power laws should not result in a perfect power law shape. To this end,
we simulate a Gaussian distribution of power law indices similar to
the one presented in Mateos et al. (\cite{mateos}), with a mean
spectral slope of 1.9 and an intrinsic dispersion of 0.23. We found
that the deviation from a power law shape for the resulting averaged
spectrum is too small to account for the observed residuals above 5
keV, thus supporting our refection component hypothesis.\\

We also attempted to fit the broad residuals by using a relativistic
line model ({\tt laor} model, Laor \cite{laor}). Leaving the line
energy to vary results in an unreasonable value of $\sim$~8~keV so,
after trying several values between 6.4 and 6.9~keV, we decided to fix
it to 6.4~keV. We found that the fit does not depend on the emissivity
index nor the inner or outer radii. The only relevant parameter
appears to be the inclination angle, but again, leaving it free to
vary results in an unphysical large value ($>$~80~$\degree$) for type
1 AGN. Requiring such a large value for the inclination angle implies
that the relativistic line is trying to fit the continuum as well as
the broad residuals. We therefore fixed the inclination angle to
30$\degree$, a typical value for type 1 AGN (Antonucci \cite{anto},
Nandra et al. \cite{nandra}, Guainazzi et al. \cite{gua}). We obtained
in this way a barely acceptable fit ($\chi^{2}$/d.o.f = 26/17),
significantly worse than the neutral reflection model.\\

\begin{figure}
\centering
\includegraphics[width=6cm,angle=-90]{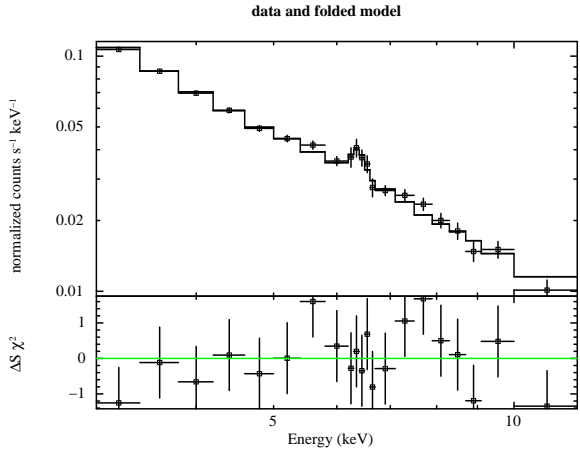}
\caption{
Spectral fit corresponding to: baseline model + pexrav.}
\label{fit}
\end{figure}

The best-fit parameters corresponding to the models that significantly
improve the ``baseline model'' fit are shown in Table
~\ref{models}. The fit corresponding to the ``baseline model'' plus
neutral reflection, which seems to be the only physically plausible case,
is shown in Fig.~\ref{fit}.\\ 

Adding a relativistic line to the ``baseline model'' plus cold
reflected component does not improve the fit. We computed an upper
limit for the EW of any broad relativistic line (for a Laor line
centered at 6.4~keV) of $<$ 400 eV at 3$\sigma$ confidence level, a
value significantly below the 560 eV reported in S05. The reasons for
these differing results could be as follows:\\

-- Differences in the sample characteristics. The sample studied in
S05 is composed of faint sources (average flux $\sim$ 10$^{-14}$ erg cm$^{-2}$ s$^{-1}$) whereas ours is composed of medium
flux sources (average flux $\sim$
5$\times$10$^{-14}$ erg cm$^{-2}$ s$^{-1}$ ). The S05 X-ray spectra are
of overall better quality than ours, having more than 200 counts in the
0.2-10 keV band, but their sample only comprizes 53 type 1 AGN whereas
ours is composed of more than 600 sources. The sources in S05
are of course more distant and luminous on average, so we are presumably 
dealing with different source populations. Higher-luminosity sources 
would be expected to have lower EW, however, due to the Iwasawa-Taniguchi 
effect (see 4.1). \\

-- Differences in the stacking method: The method we have developed
takes into account all the possible contributions to the underlying
continuum. As can be seen in the simulations (see Appendix A), broad
residuals can appear due to dispersion in the underlying
continuum. The subtraction of the continuum has been dealt with in a
less detailed way in S05, and this could result in an overestimation
of the broad line EW. Besides, using grouped spectra, as in S05, can
introduce features such as a broad red tail in an emission line as
showed in Yaqoob et al. (\cite{yaqoob}).  \\
\subsection{Sub-samples}
\begin{figure}[t]
\centering
\includegraphics[width=6cm,angle=-90]{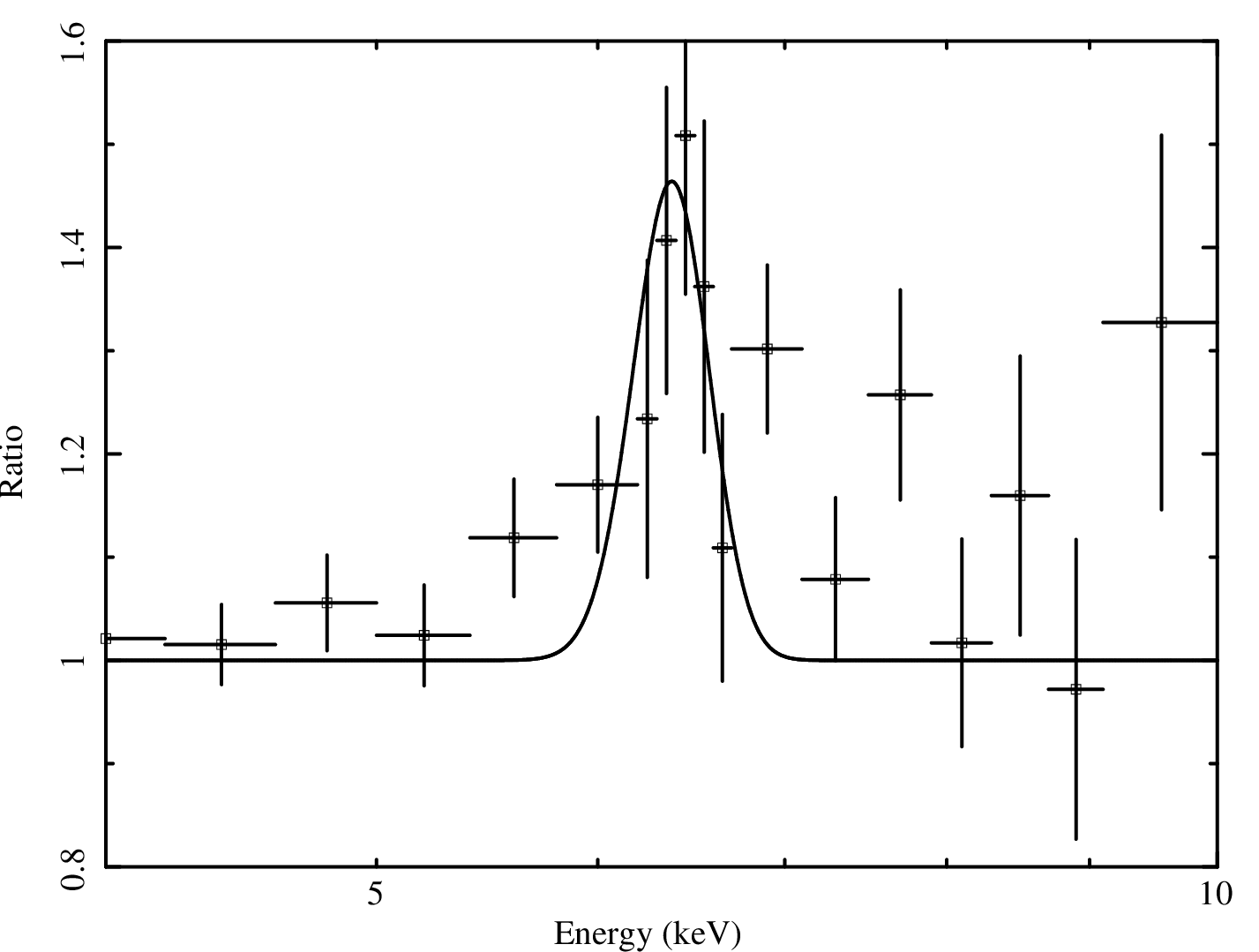}
\includegraphics[width=6cm,angle=-90]{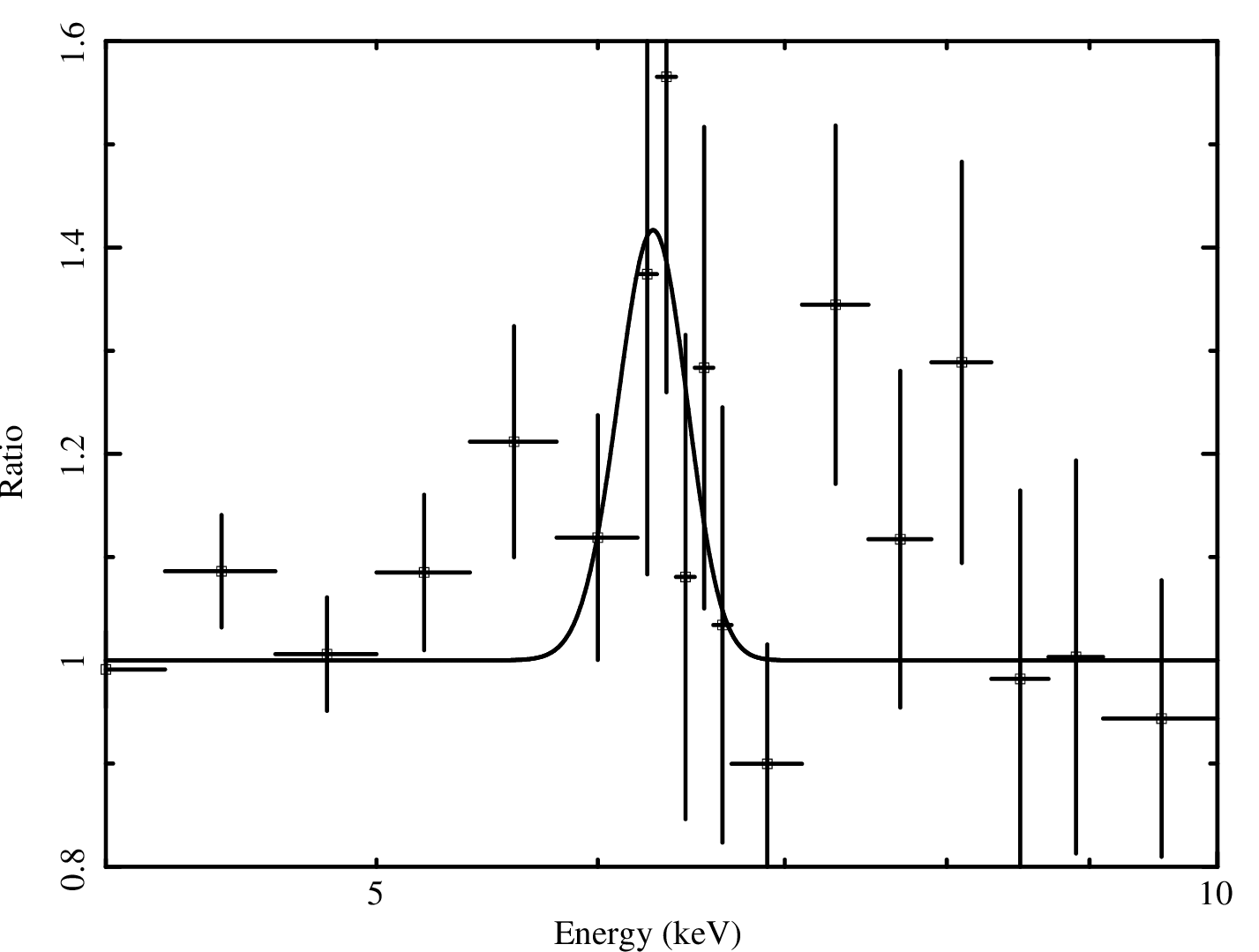}
\includegraphics[width=6cm,angle=-90]{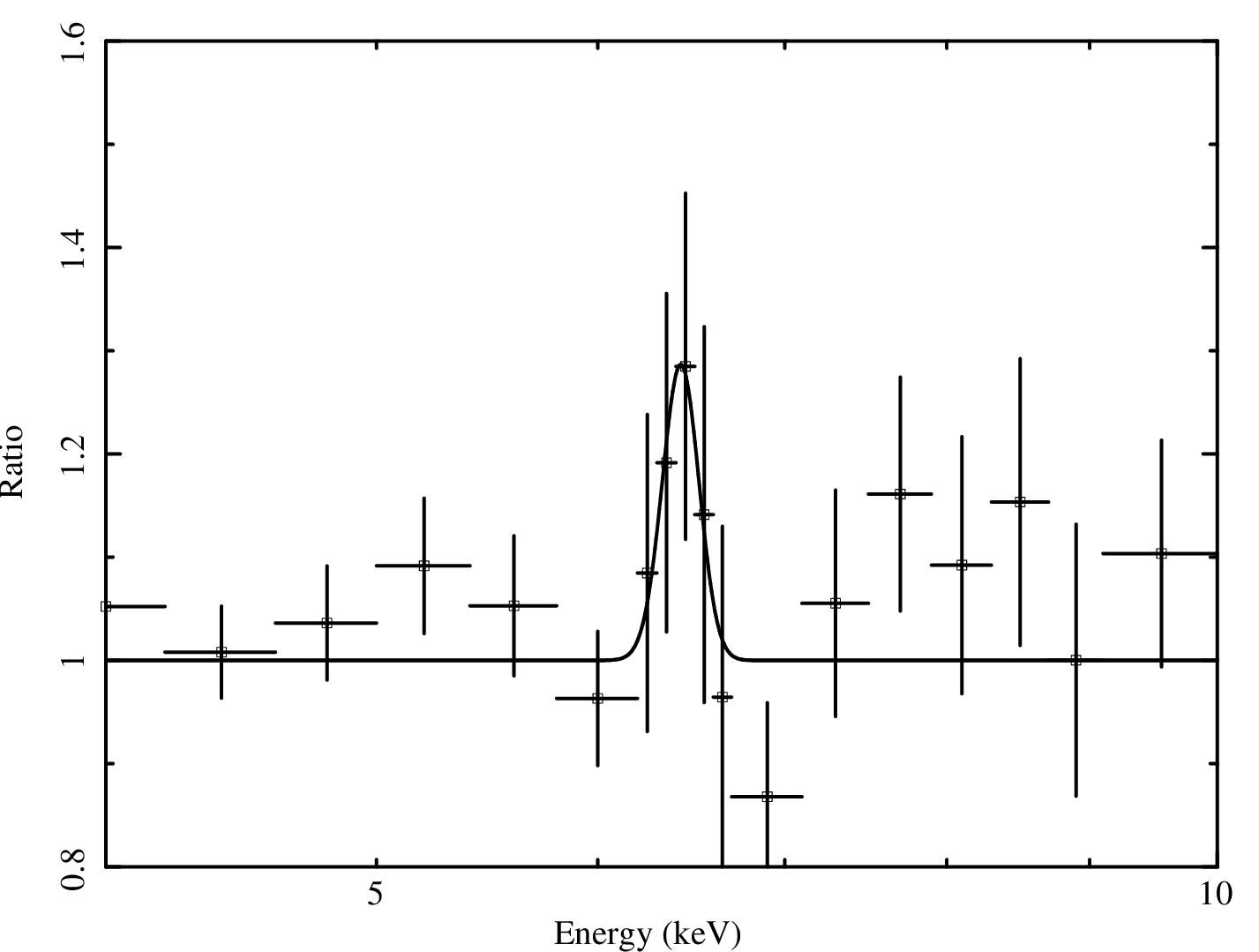}
\caption{
Soft (0.5-2.0~keV) Luminosity samples. {\it Top panel}: Averaged spectrum with Gaussian fit for the low luminosity sample.{\it Central panel}: Medium Luminosity sample.{\it Bottom panel}: High Luminosity sample.}
\label{lumin}
\end{figure}
\begin{table}
\centering
\caption[]{Luminosity sub-samples.}
$$
\begin{array}{c c} 
\hline
{\rm Luminosity\;Range} & {\rm Narrow\;FeK\alpha\;EW} \\
{\rm (erg\,s^{-1})} & {\rm (eV)} \\
\hline 
\hline 
{\rm 1.0\times10^{42}-1.8\times10^{44}} & {\rm 190^{+40}_{-50}}\\
{\rm 1.8\times10^{44}-6.1\times10^{44}} & {\rm 150^{+70}_{-100}}\\
{\rm 6.1\times10^{44}-6.6\times10^{46}} & {\rm 50^{+50}_{-40}} \\ 
\noalign{\smallskip} 
\hline
\end{array}
$$
\label{lumintable}
\end{table}

Given the relatively large number of counts in the final averaged spectrum,
we divided the sample into three different bins containing the same
number of counts, in order to check for possible averaged spectral
variations as a function of redshift, flux or luminosity. We did not
detect any significant variation with redshift nor with flux, although
we marginally detected a correlation of the narrow emission line EW with
Luminosity. As can be seen from Fig.~\ref{lumin} and Table
~\ref{lumintable}, the narrow line EW decreases as the luminosity increases, a
behaviour previously reported as the Iwasawa-Taniguchi or X-ray
Baldwin effect (Iwasawa \& Taniguchi, \cite{iwa}), but the errors are
too large to extract conclusive results. The results that are
presented both in the Fig.\ref{lumin} and Table ~\ref{lumintable}
correspond to the luminosity sub-samples selected in the soft band
(0.5-2.0~keV) but we also observed this behavior when selecting in the
hard (2.0-10.0~keV) and XID (0.5-4.5~keV) bands. Note that the
tabulated values are very likely higher than the actual values since
we are not fitting the reflection component in this case.\\

A similar result was found in K08. They found that low luminosity
  sources (0.5-10 keV luminosity $<\,10^{44}\,erg\,s^{-1}$) show a
  prominent emission line at $\sim$ 6.6 keV with an EW $\sim$ 2 keV
  whereas high luminosity sources do not show any significant excess
  emission in the 4 to 9 keV range.

%%%%% Resultados
\section{Conclusions}

We developed a new method to obtain an X-ray averaged spectrum for AGN
and applied it to the identified type 1 AGN in the XWAS and AXIS
samples. We extensively checked this method and performed several
tests by using simulated spectra. We also used a large set of
simulated spectra to construct an underlying continuum against which
the average spectrum is compared. We measured a spectral slope of
$\Gamma=1.9\pm0.02$ in the average spectrum. We significantly
detect a narrow line at 6.36$\pm$0.05 keV corresponding to the Fe
K$\alpha$ emission line emitted from neutral (or low-ionization)
material, i.e. far from the central engine, with an EW = 90 $\pm$ 30
eV. This result is in agreement with several studies of local AGN
showing that a narrow Fe K$\alpha$ emission line, displaying similar
characteristics, is almost universally present in the X-ray spectra 
of nearby AGN.\\

We do not significantly detect any significant broad relativistic
emission line, although some positive residuals can be seen in the 5
to 10~keV region. We find that the continuum shape is best represented
by reflection from neutral or low-ionization material rather than by a
relativistic iron line, also consistent with studies of local AGN. The
presence of a mixture of low EW relativistic lines is not completely
excluded, which might suggest a small amount of emitting material
close to the ISCO. We obtained an upper limit for the corresponding EW
of 400 eV at a 3$\sigma$ significance level, a value significantly
lower than the one presented in Streblyanska et
al. (\cite{streb}). This difference can be attributed either to our
more sophisticated treatment of the underlying continuum or to a
difference in the source populations themselves. To test these
hypotheses, we are increasing the number of sources in our sample and
constructing a large sample of type 2 AGN, by adding data from both
deeper and shallower samples. Results from these new samples will be
presented in a forthcoming paper.\\

Our results are very important for one of the central scientific goals
of the XEUS\footnote{http://www.rssd.esa.int/XEUS} mission, which is
to detect growing AGN at significant redshifts (z $\sim$
5-10). Although XEUS is being designed to have the sensitivity to
detect these sources, their identification will not be a trivial task,
even in the 2020's with an expected plethora of large observatories at
all wavelengths. However, our result that a narrow Fe line is common
in type 1 AGN in the distant Universe, with an EW $\sim$ 100 eV,
confirms the expectation that redshifts can be obtained for these
early AGN from their X -ray spectrum directly.

\begin{acknowledgements}
Partial financial support for this work was provided by the Spanish
Ministerio de Educaci\'{o}n y Ciencia under project
ESP2006$-$13608. AC acknowledges financial support from a Spanish
Ministerio de Educaci\'{o}n y Ciencia fellowship and also from the
MIUR and The Italian Space Agency (ASI) grants PRIN$-$MUR
2006$-$02$-$5203 and n. I/088/06/0. MJP, SM, JAT \& MGW acknowledge
support from the UK STFC research council. MC is supported by the
Deutsches Zentrumfuer Luft-und Raumfahrl (DLR) GmbH under contract
No. FKZ50 OR 0404. We thank the referee, Bev Wills, for useful
suggestions.

\end{acknowledgements}

\begin{appendix}
\section{Simulations}
In order to distinguish between real spectral features and artifacts
from both the averaging process and/or the individual spectral shape
around a possible emission line, we carried out extensive quality
tests by using simulated spectra. Our strategy can be summarized as
follows: We simulate N times each source, including Poisson counting
noise, keeping the same 2-8~keV observed flux and exposure time as in
the real spectrum along with the same calibrations files. We then
applied to the simulated spectra exactly the same averaging method as
for the real spectra. We also tested how the spectral model, noise and
number of simulations affected the final averaged simulated spectra.\\

\begin{figure}
\centering
\includegraphics[width=6cm,angle=-90]{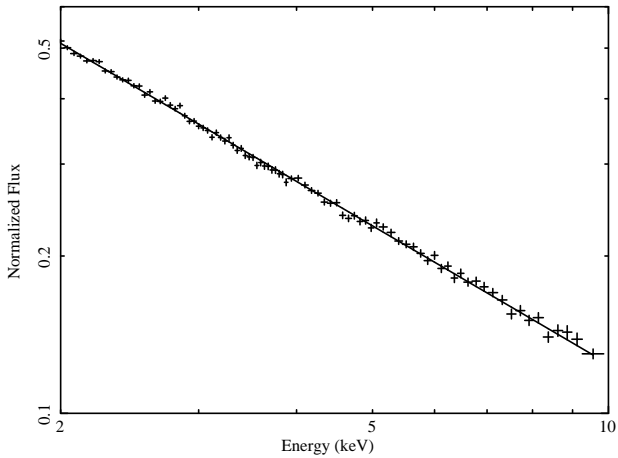}
\caption{
{\bf Model: po.} Simulated average spectrum corresponding to 10 simulated spectrum for each real spectrum and a simple power law model, $\Gamma$=1.9.}
\label{po}
\end{figure}

First, we performed 10 simulations for each source, in order to reduce
both noise and spurious features, using a simple power law model with
$\Gamma$ = 1.9 constant ({\tt XSPEC} model: po). The resulting averaged
spectrum is shown in Fig.~\ref{po}. We fitted a power law between 2 and
10~keV obtaining a $\Gamma$(2-10keV)=1.89$\pm$0.01 and we did not
detect any significant features. We also checked for the effect of not
including Poisson noise and increasing the fluxes, obtaining the same
result but with smaller errors.\\

\begin{figure}
\centering
\includegraphics[width=6cm,angle=-90]{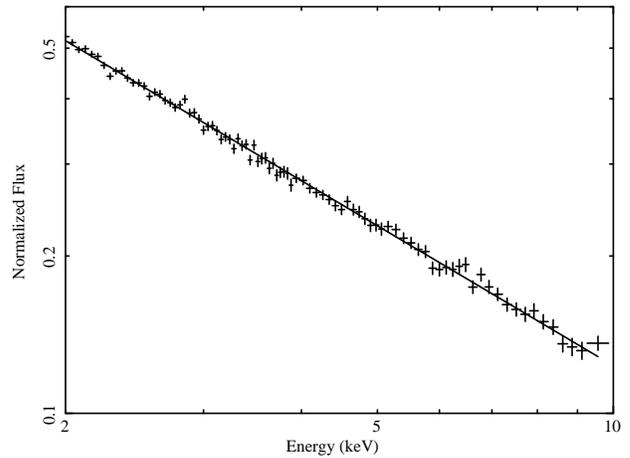}
\caption{
{\bf Model: pha*po.} Simulated average spectrum corresponding to 10 simulated spectrum for each real spectrum and a power law ($\Gamma$=1.9) plus Galactic absorption model.}
\label{phapo}
\end{figure}

Our method includes the correction for the Galactic absorption for
each source, so we needed to check how this correction affects the
average spectrum. We then simulated 10 times each source with a power
law model ($\Gamma$ = 1.9) modified by Galactic absorption ({\tt
  XSPEC} model: pha*po). The resulting average spectrum is shown in
Fig.~\ref{phapo}. We again measured $\Gamma$(2-10~keV)=1.89$\pm$0.01
and did not detect any significant features. We can conclude that the
correction for the Galactic absorption does not affect the results
significantly. \\

\begin{figure}
\centering
\includegraphics[width=6cm,angle=-90]{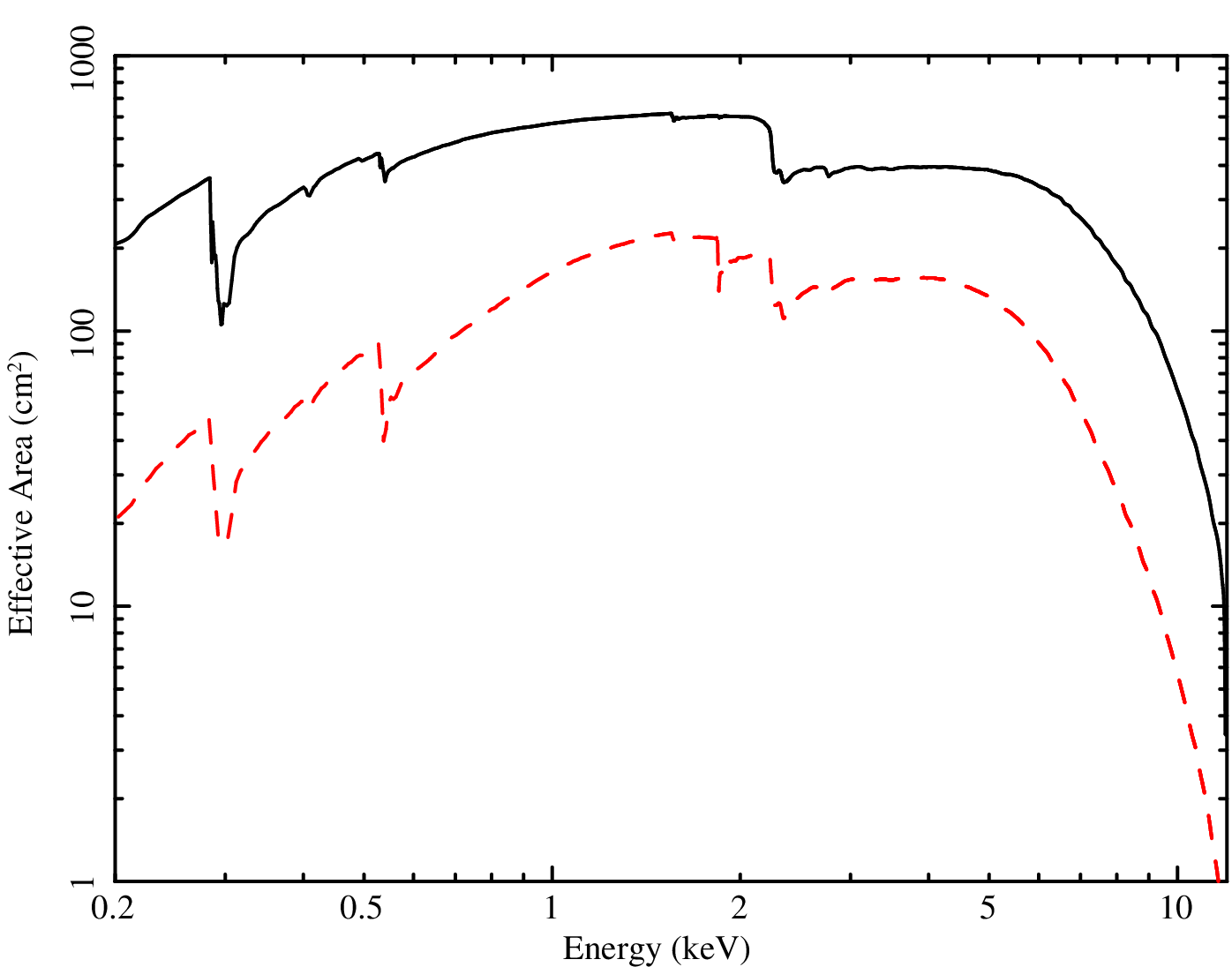}
\includegraphics[width=6cm,angle=-90]{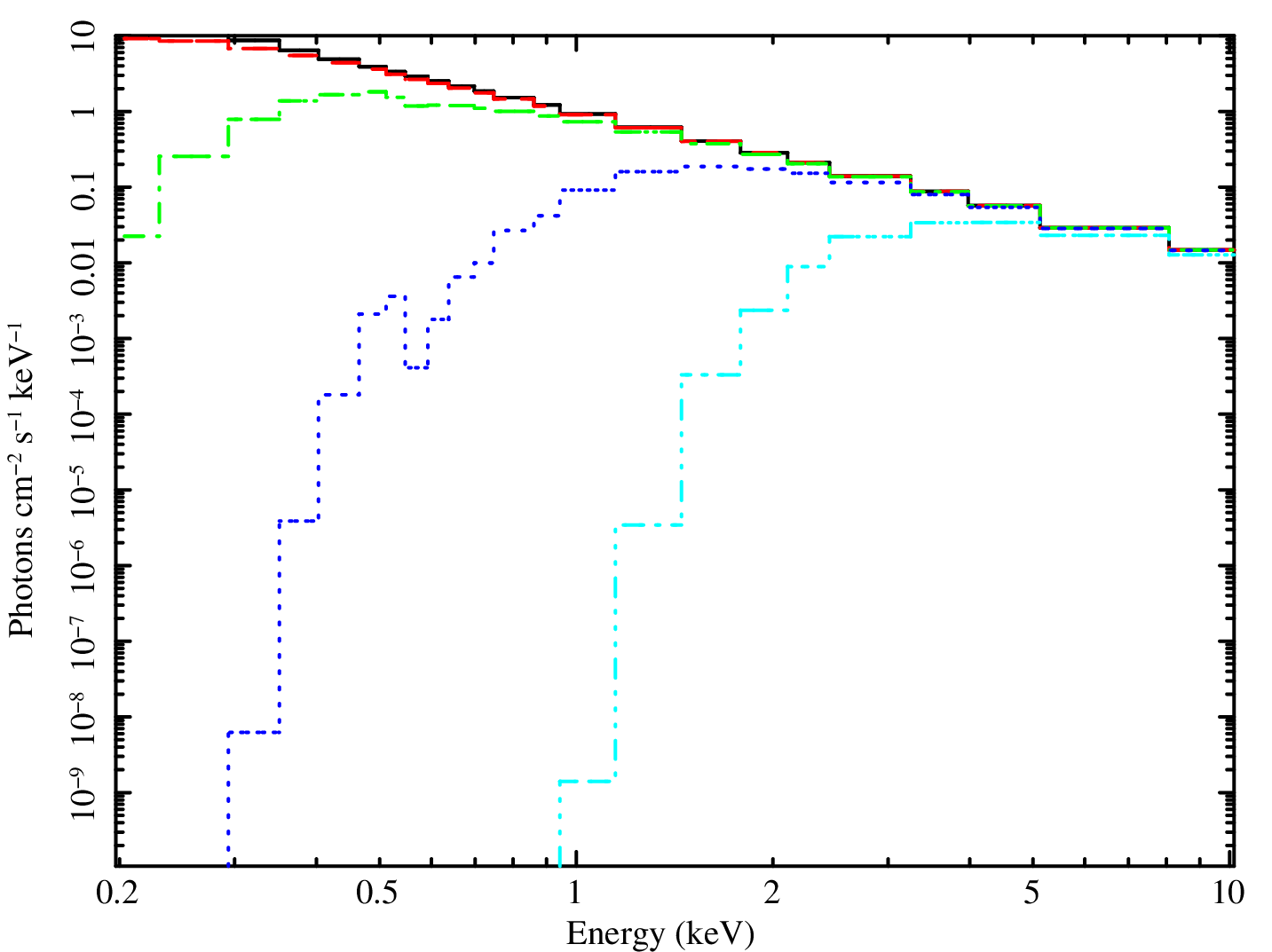}
\caption{{\it Top panel:} Effective area corresponding to an individual observation for pn (solid line) and MOS cameras (dashed line). {\it Bottom panel:} pha*po models for different amounts of absorption: no absorption (solid line), 10$^{20}$cm$^{-2}$ (dashed line), 10$^{21}$cm$^{-2}$ (dash-dotted line), 10$^{22}$cm$^{-2}$ (dotted line), 10$^{23}$cm$^{-2}$ (dash-dot-dot-dotted line).}
\label{examples}
\end{figure}

Once we had tested that the method and the counting statistics do not
significantly affect the results, we had to test how narrower spectral
features propagate through our averaging and analytic procedures. It
is well known that intrinsic absorption, along with warm absorption
and reflection components, can significantly affect the continuum
below the Fe emission lines. Unfortunately, the spectral quality of
most of our spectra did not allow us to fit complex models and
subtract them in order to obtain a pure emission line
spectrum. Besides, the unfolding procedure itself can also affect the
spectra due to the shape of the effective area (see
Fig.~\ref{examples}), although it has been tested that this does not
significantly affect the spectral shape above 3~keV.\\

\begin{figure}
\centering
\includegraphics[width=6cm,angle=-90]{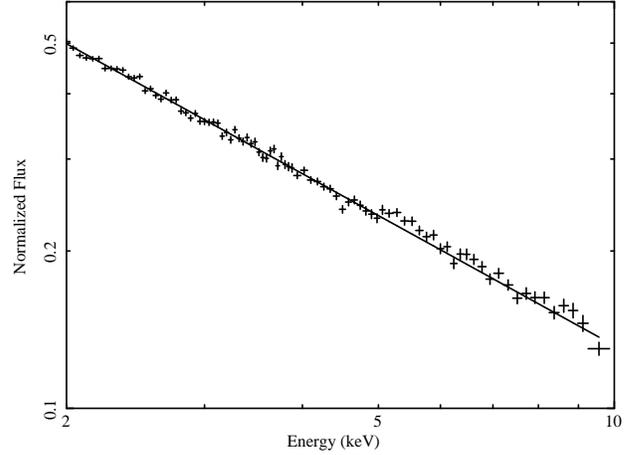}
\caption{
{\bf Model: pha*zpha*po.} Simulated average spectrum corresponding to 10 simulated spectrum for each real spectrum and a power law ($\Gamma$=1.9) plus Galactic absorption and intrinsic absorption model.}
\label{phazphapo}
\end{figure}

In order to quantify the effect of intrinsic absorption, we used the
``best fit model'' we obtained for the real spectra (an absorbed power
law {\tt XSPEC} model: pha*zpha*po) and simulated 10 times each
source, Fig.~\ref{phazphapo}. We can see broad features above 4~keV,
and we measured a $\Gamma$(2-10~keV)=1.81$\pm$0.01, thus the average
spectrum is affected by the intrinsic absorption above 3~keV and it
has to be accounted for very carefully. As an example, in
Fig.~\ref{examples} we show the variation in the spectral shape for a
power law ($\Gamma$ = 1.9) as the amount of absorption increases from
10$^{20}$ to 10$^{23}$ cm$^{-2}$ (at a typical spectral resolution),
although only 5$\%$ of the type 1 AGN in our sample show an intrinsic
absorption above 10$^{22}$ cm$^{-2}$.\\

\begin{figure}
\centering
\includegraphics[width=6cm,angle=-90]{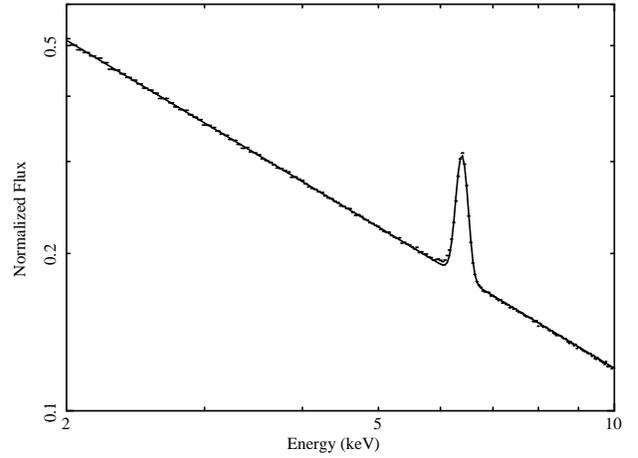}
\caption{{\bf Model: po+gaus}: Simulations without noise corresponding to E=6.4~keV, $\sigma$ =0 and EW=200 eV, rest-frame values.}
\label{pogaus}
\end{figure}

To test how narrow features are treated by our method, we simulated a
power law plus an intrinsically unresolved emission line ($\sigma$ =
0). To avoid the contribution from the noise, we did not include
Poisson noise this time and increased the source flux by a large
factor. We were not interested in effects introduced at these low
fluxes or counting statistics but only in the effects due to the
process and the redshift distribution over a narrow feature. We
simulated once each source, using a power law plus a Gaussian emission
line ({\tt XSPEC} model: po+gaus) with the following parameters:
$\Gamma$=1.9, E = 6.4~keV, $\sigma$ = 0 and EW = 200 eV, rest-frame
values. The result is presented in Fig.~\ref{pogaus}. We can see how
the process and the EPIC spectral resolution widen the line giving a
value of $\sigma\simeq$100 eV, but we recovered the initial values for
the remaining parameters ($\Gamma$, E and EW). We therefore see that
our averaging method introduces a $\sim$ 100 eV broadening on narrow
features that we subtract in quadrature from fitted values.\\

\begin{figure}
\centering
\includegraphics[width=6cm,angle=-90]{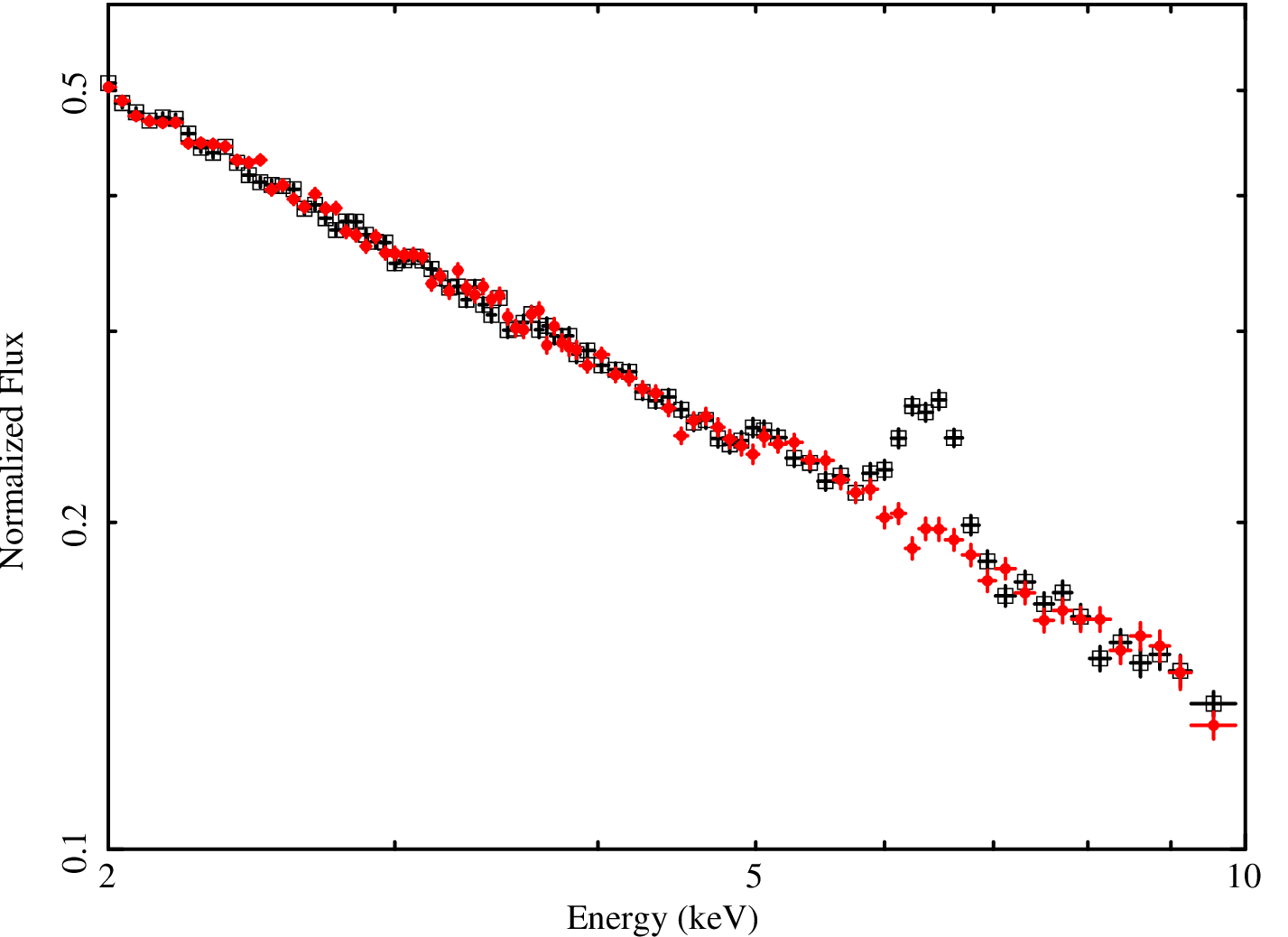}
\includegraphics[width=6cm,angle=-90]{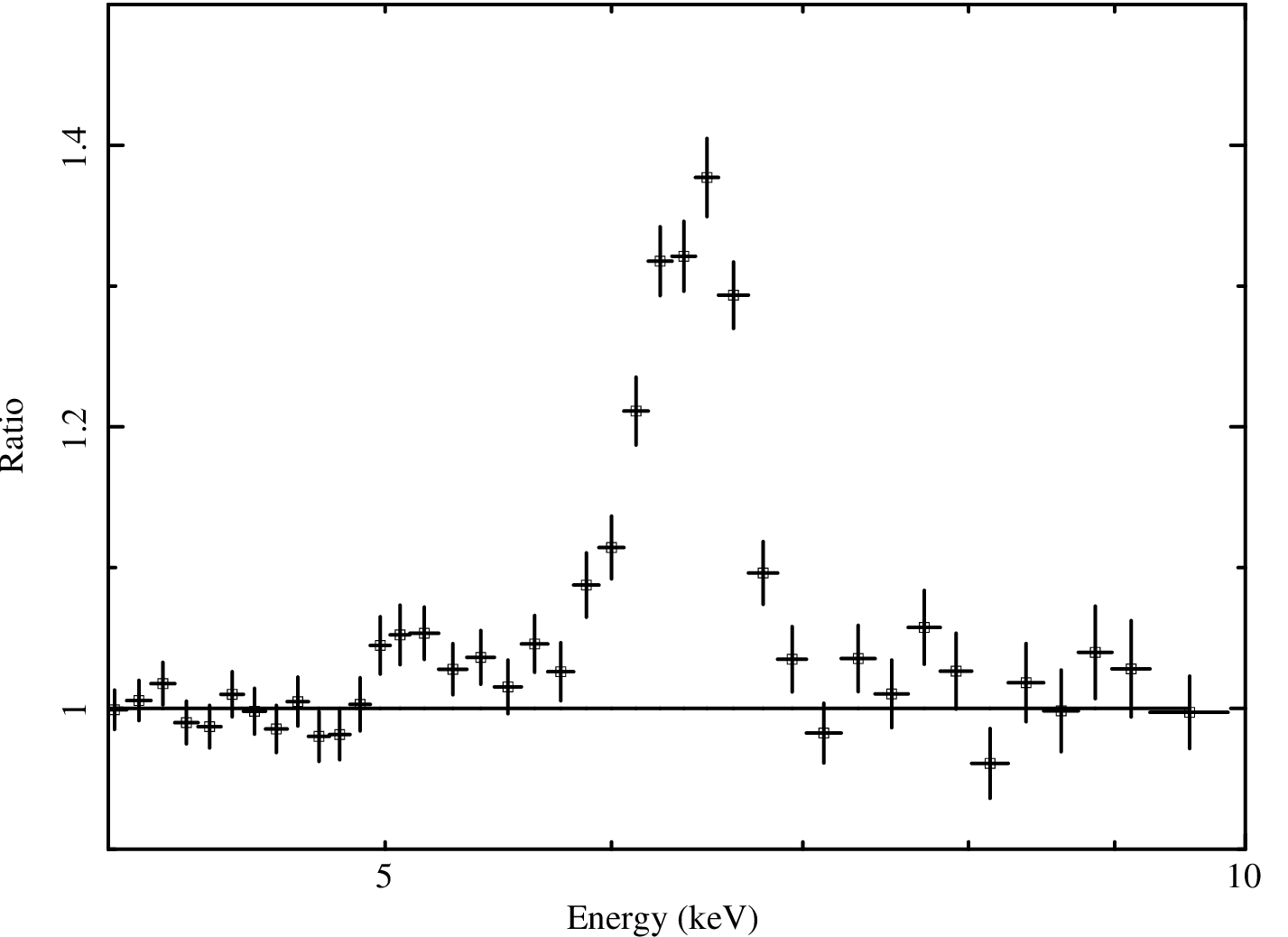}
\includegraphics[width=6cm,angle=-90]{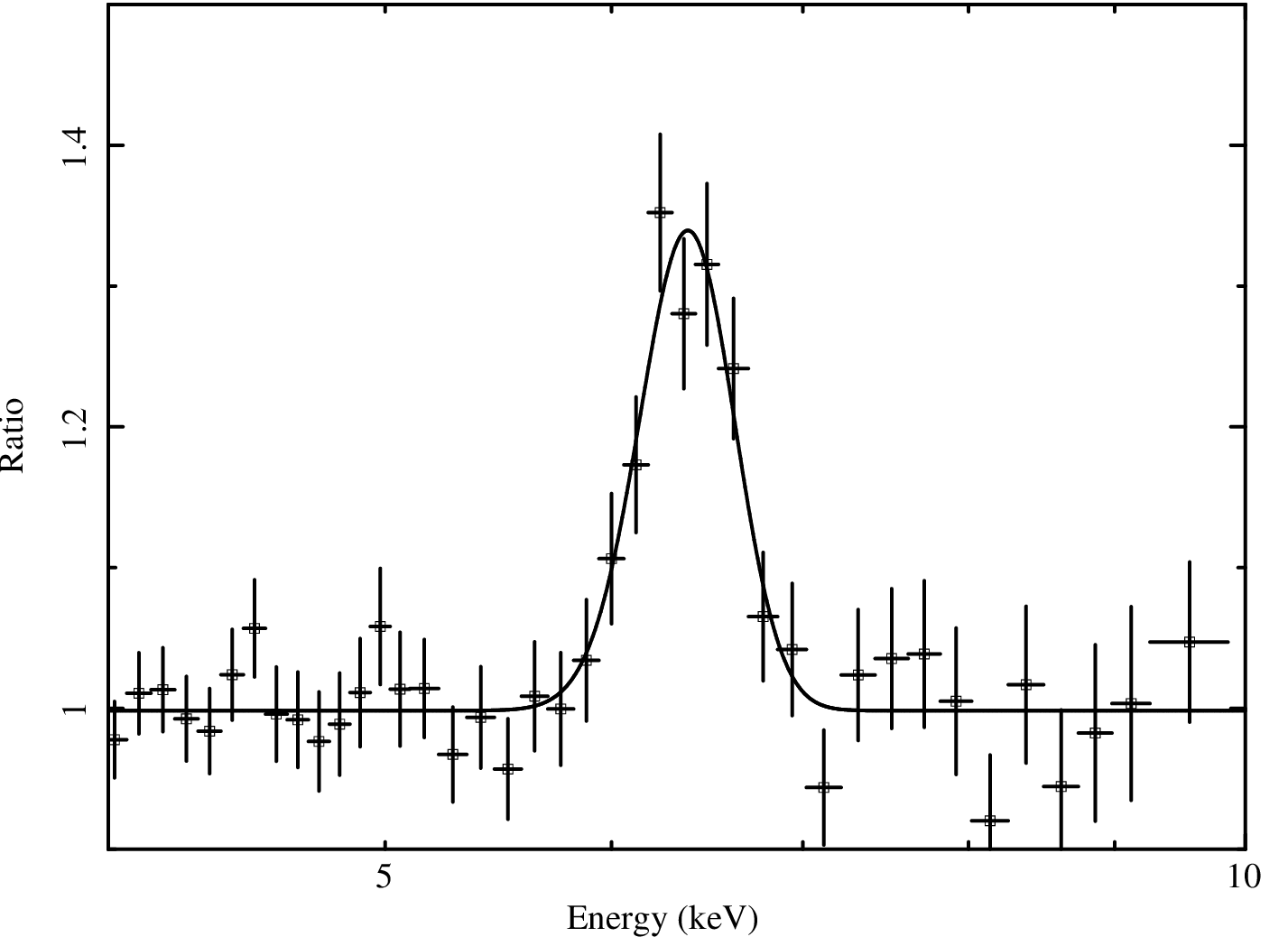}
\caption{ {\it Top panel:} {\bf Model: \bf pha*zpha*(po+gaus)}:
  Simulated average spectrum corresponding to 10 simulated spectra
  for each real spectrum (squares) and the ``best-fit model''
  continuum (circles). {\it Central Panel:} Simulations
  pha*zpha*(po+gaus) to power law ratio. {\it Bottom Panel:}
  Simulations pha*zpha*(po+gaus) to ``best-fit model'' continuum
  ratio.}
\label{phazphapogaus}
\end{figure}
\begin{figure}
\centering
\includegraphics[width=6cm,angle=-90]{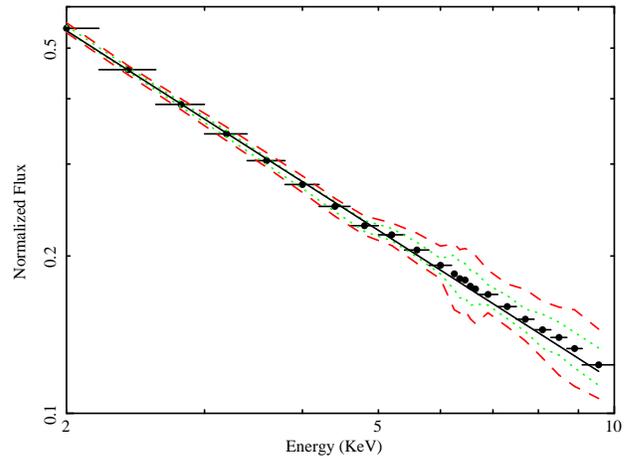}
\caption{Average continuum with 100 simulations (circles) along with the
  power law fit (2-10~keV) and the 1$\sigma$ (dotted line) and
  2$\sigma$ (dashed line) confidence limits.}
\label{contours}
\end{figure}

At this point we still needed to quantify how absorption could
affect the shape of an emission line. To achieve this, we added a
Gaussian emission line to the ``best fit model'' and ran the
simulations using this model ({\tt XSPEC} model:
pha*zpha*(po+gaus)). Fig.~\ref{phazphapogaus} (top) shows the
simulations for a Gaussian line centered at E = 6.4~keV with
$\sigma\simeq$ 100 eV and EW = 100 eV (rest-frame values) along with
the simulations for the ``best-fit model'' ({\tt XSPEC} model:
pha*zpha*po). We can see that the line shape appears distorted towards
low energies mimicking a broad tail, but this feature is also present
in the pha*zpha*po simulations. If we fit a power law in the 2-10~keV
range, excluding the emission line region (4-7~keV), and compute the
ratio between the pha*zpha*(po+gaus) model and this power law, we
obtain the ratio shown in Fig.\ref{phazphapogaus} (central). The line
shape is clearly distorted and some residuals below 6~keV can be
seen. Notwithstanding, if we use the simulations without a line as an
underlying continuum to obtain the ratio, Fig.~\ref{phazphapogaus}
(bottom), we find that the process slightly widens the line due to
the unfolding and the different energy resolutions at different
energies, but it does not vary the line shape nor its EW, recovering
the input parameters within errors. We also tested if changing the
line width, EW and central energy did not affect significantly our
conclusions, always recovering the input values within errors when
using the pha*zpha*po simulations as the underlying continuum.\\

To minimize the distortion introduced by the averaging method and the
intrinsic absorption for a possible emission line, we decide to use
this simulated continuum as an underlying continuum for our real
average spectrum. Instead of simulating each source 10 times we used
100 simulations per real source, so we were able to construct 100
simulated continua. From these 100 simulated continua we could
quantify the significance of any excursion over this continuum by
constructing confidence limits, i.e. flux intervals per each energy
grid in which a number of simulations is contained (see
Fig.~\ref{contours}). For the 100-simulations continuum we fitted a
power law between 2 and 10~keV obtaining $\Gamma$=1.96$\pm0.01$, close
to the mean value for the real sample. Although the continuum shape
does not show prominent features the confidence contours display an
increasing dispersion towards high energies and around the position
where the Fe K$\alpha$ line is expected to fall. Therefore, this must
be taken into account when fitting a possible broad line as it could
be very much affected by these continuum features.

\end{appendix}
\end{document}